# Hessian filter-assisted full diameter at half maximum (FDHM) segmentation and quantification method for optical-resolution photoacoustic microscopy

**Dong Zhang [a,b], Ran Li [c], Xin Lou [b,*] and Jianwen Luo [a,*]**

[a] Department of Biomedical Engineering, School of Medicine, Tsinghua University, Beijing, China
[b] Department of Radiology, Chinese PLA General Hospital, Beijing, China
[c] School of Basic Medical Sciences, North China University of Science and Technology, Tangshan, Hebei, China

**Abstract**

Optical-resolution photoacoustic microscopy has been validated as a high-resolution and high-sensitivity imaging modality for angiographic studies in the past decades. Quantitative vascular analysis reveals critical information of physiological changes, where vessel segmentation is the key step. In this work, we developed a Hessian filter-assisted, adaptive thresholding vessel segmentation algorithm. Its performance is validated by a digital phantom and *in vivo* images. Its capability of capturing subtle vessel changes is further tested in two longitudinal studies on vascular responses to blood pressure agents. The results are compared with the widely used Hessian filter method. In the antihypotensive case, the proposed method detected a twice larger vasoconstriction than the Hessian filter method. In the antihypertensive case, the proposed method detected a vasodilation of 18.8 %, while the Hessian filter method failed in change detection. The proposed algorithm could correct errors caused by conventional segmentation methods and improve quantitative accuracy for angiographic applications.

∗ Corresponding authors.

E-mail address: luo_jianwen@tsinghua.edu.cn (J. Luo), louxin@301hospital.com.cn (X. Lou)

## 1. Introduction

Photoacoustic imaging (PAI), a hybrid technique detecting tissue absorption properties ultrasonically, has been a rapidly developing modality for preclinical and clinical studies in the past decades. PAI enables investigation of structural and functional information at various scales ranging from organelles to organs [1]. As a main embodiment of PAI, optical-resolution photoacoustic microscopy (OR-PAM) is able to achieve high-contrast, high-resolution and high-sensitivity imaging by confocal and coaxial configuration of the tightly focused optical excitation and high frequency acoustic detection. Regarding the high absorption to the visible and near-infrared light, hemoglobin serves as an ideal endogenous contrast for OR-PAM imaging of vasculatures *in vivo*. The inherent advantages of OR-PAM and associated endogenous contrasts make it a powerful tool for angiographic studies.

Vascular features are typically accompanied with physiological processes. Multiple parameters extracted from vascular morphologies are widely used for diagnosis [2]. Diameter distribution change in the brain reflects cerebral blood flow regulations [3]; vessel area fraction (VAF) change is a general sign of vasoconstriction or vasodilation, indicating physiological responses to pressure medicines [4]. In order to acquire these parameters, an accurate vessel segmentation method is required to extract vessel networks from gray-scaled vascular images.

In previous studies, simple thresholding approaches have been applied to segment microvasculature, among which the Otsu's thresholding (OT) shows promising results [5]. Although this single-value thresholding approach is likely to either miss small vessels or overly segment large ones, it is still widely in use because of its simplicity. Numerous advanced vessel segmentation techniques have been developed, which can mainly be classified into two categories, i.e., tracking methods and filter-based methods. The tracking methods trace the vessels iteratively by starting from a pre-selected vessel point, and locating the next until the entire network is extracted. The vessels are segmented during the tracking process. The tracking methods perform well on detecting large vessels, but could fail on small vessels due to their weak signal and small scales [6]. The filter-based methods generally enhance the vascular contrast. It extracts vessel structures by applying a filter, which increases the vessel contrast when the vascular size and orientations are matched by the filter. Hessian filter (HF) is still one of the mostly used techniques since its report 20 years ago, for its high sensitivity to tubular structures and low computational cost [7]. However, the vessel distortions brought along require further explorations on this technique [8]. Yang et al. and Zhao et al. developed a Hessian filter-based method which achieved better vessel enhancement on PAM images [9, 10]. Khan et al. reported an approach by fusing two segmented results of different filter scales to mitigate the distortions [11]. Although these studies have pushed forward the limit of the HF method, the manually selected scale ranges would in any way introduce errors to the results.

Different from large vessels in OR-PAM images, small ones generally possess weak signal, which is difficult to extract. Also, user-selected parameters in the HF algorithm, like scale ranges or thresholds, may further cause errors because of different user experience. In this work, a robust vessel segmentation algorithm of high accuracy for PAM vascular images is proposed. It applies HF for small vessel extraction and fuses the results with large vessels and crossing points segmented by OT. By upward region growing (URG), the lost connectivity caused by HF is restored; and by taking the full diameter at half maximum (FDHM) pixel by pixel along the skeleton of the vessel branches, the distorted vessel sizes are corrected. The performance of the algorithm is validated by a custom-established 3D digital phantom and *in vivo* mouse cortex and ear vasculature. The algorithm is further tested with two longitudinal studies on vascular responses to a vasoconstricting and a vasodilating drug.

## 2. Materials and Methods

### 2.1. Algorithm workflow

The overall flowchart and corresponding intermediate example images of the proposed algorithm is shown in **Figure 1.** The segmentation algorithm consists of four steps: preprocessing, vessel enhancing, thresholding and segmenting, as shown in **Figure 1 (a)**.

**Preprocessing**. In the preprocessing step, a mean filter and a median filter are applied for denoising, followed by a contrast enhancing process. In a maximum amplitude projection (MAP) vessel image, system noise can be measured by recording signal without imaging any target. The system noise is subtracted and the top 1% pixel values are saturated to increase the contrast. A high-frequency emphasis (HFE) filtering is then performed, which emphasizes the high frequency components, referring to small vessels, while maintaining the rest frequency components of the image. The emphasized image is subsequently processed by a contrast limit adaptive histogram equalization (CLAHE) operator, which improves the local contrast to further differentiate the small vessels from background [12]. Notably, the background noise is also enhanced by CLAHE, as shown in **Figure 1(b)**, but would be suppressed in the next steps.

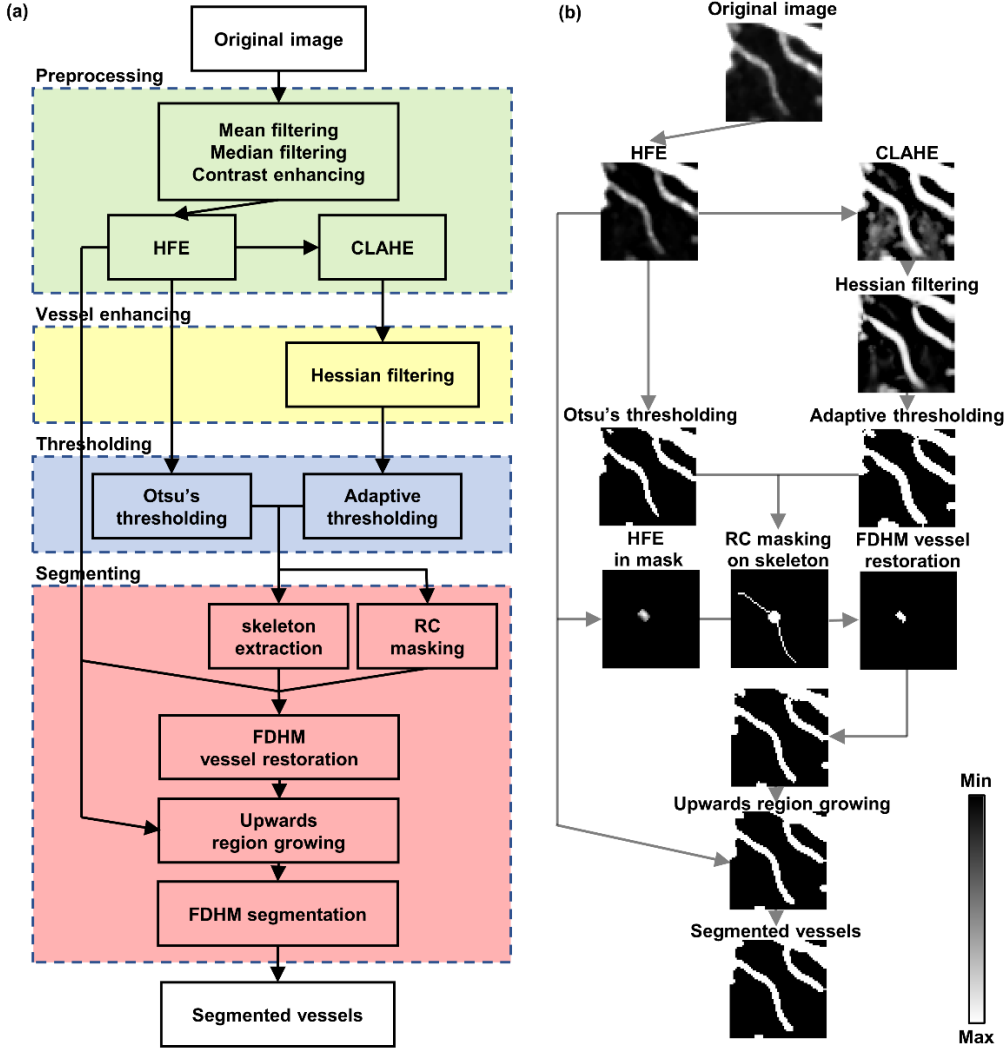

**Figure 1** The flowchart and corresponding intermediate example images of each step of the proposed algorithm. (a) Flowchart of the proposed algorithm. (b) Corresponding intermediate example images after each step. HFE, high frequency emphasis; CLAHE, contrast limit adaptive histogram equalization; RC, radius circular; FDHM, full diameter at half maximum.

**Vessel enhancing.** The vessel enhancing step is achieved by HF [7]. In this step, the vessels of an image are probed by a Hessian matrix and the tubular structures in the image are extracted by calculating the eigenvalues of the matrix. The Hessian matrix at pixel $(x, y)$ is defined as:

$$H = \begin{bmatrix} I_{xx}(x,y,\sigma) & I_{xy}(x,y,\sigma) \\ I_{yx}(x,y,\sigma) & I_{yy}(x,y,\sigma) \end{bmatrix} \quad (1)$$

with $I_{xy}(x, y, \sigma)$ written in a well-posed form according to the concept of linear scale space theory [7]:

$$I_{xy}(x, y, \sigma) = I(x, y) * \sigma^2 G_{xy}(x, y, \sigma) \quad (2)$$

where $I(x, y)$ is the pixel value, $G_{xy}(x, y, \sigma)$ is the second-order partial derivative of the multi-scale Gaussian kernel with the size determined by $\sigma$. From the Hessian matrix, the eigenvalues $\lambda_1$ and $\lambda_2$ are calculated. To enhance the tubular structures in an image, two numerical operators are designed based on the eigenvalues:

$$R_B = \frac{|\lambda_1|}{|\lambda_2|} \quad (3)$$

$$S = \sqrt{\lambda_1^2 + \lambda_2^2} \quad (4)$$

When a pixel belongs to a vessel, the two eigenvalues hold the relation of $|\lambda_1| \approx 0$ and $|\lambda_1| \ll |\lambda_2|$, making $R_B$ very small and $S$ large. If the pixel falls in background, the two eigenvalues are both very small and thus $R_B$ is large while $S$ is small. The enhanced image is therefore defined as:

$$f(x, y, \sigma) = \begin{cases} 0 & \lambda_2 > 0 \\ \exp\left(-\frac{R_B^2}{2\beta^2}\right)\left(1 - \exp\left(-\frac{S^2}{2c^2}\right)\right) \end{cases} \quad (5)$$

where $\beta$ and $c$ are two weights to adjust the sensitivity to the two operators. The tubular structures are measured at different scales and only when the scale of the kernel, $\sigma$, matches the size of the vessel, $f(x, y, \sigma)$ reaches its maximum. So, the enhanced vessels are obtained by:

$$f(x, y) = \max_{\sigma_{min} < \sigma < \sigma_{max}} f(x, y, \sigma) \quad (6)$$

The HF extracts vessels and suppresses non-tubular background signal. However, it also introduces vessel distortions due to the user-selected kernel scales and fails to detect the branching and crossing points because of its sensitivity solely to tubular structures. Two types of distortions are brought into the process, i.e., the vessel size distortion and the disconnectivity of adjacent vessels. As for the vessel size distortion, when the filter scale range covers all vascular sizes, the smaller vessels are dilated by large filters; when the scale range is not large enough, the filters only enhance the edges of large vessels while suppressing the signal inside, forming 'hollow' large vessels [8]. The shape of the second-order partial derivative of the Gaussian kernel inherently determines that it enhances the vascular pixels and suppresses the pixels right next to that vessel, leading to vessel disconnectivity [7]. In addition, the branching and crossing points are less likely to be extracted because they are not tubular in shape, to which the HF is not sensitive. These distortions from HF are the key reasons for its inaccuracy in quantification, and will be corrected in the following steps.

**Thresholding.** In the thresholding step, an adaptive thresholding strategy converts the Hessian filtered image into binary and the OT extracts the large vessels as well as branching and crossing points. The adaptive thresholding is implemented with two Gaussian filters with the sizes of 3 pixels and 11 pixels, both convolving with the Hessian filtered image to form two smoothed maps. At each pixel, the smaller value of the two maps serves as the corresponding threshold [13]. The idea behind this strategy is to set a lower threshold for small vessels and a higher one for large vessels, to practically preserves small vessels and avoid over segmentation on large ones. The OT is simultaneously applied on the HFE filtered image to acquire a binary map of high pixel-value components, which are large vessels as well as branching and crossing points. Small and large vessels, branching and crossing points are all extracted by fusing the two binary maps. Also, the 'hollow' large vessels are filled up. Note that, the vascular size distortion introduced by the HF and adaptive thresholding step will be corrected in the following steps, so the choice of the two Gaussian filter sizes is not necessarily strict as long as the target vessels are preserved.

**Segmenting**. In the segmenting step, the lost connectivity is restored with FDHM vessel restoration followed by URG, and the vessel size distortion is finally corrected by FDHM segmentation.

In the FDHM vessel restoration step, the vascular skeleton and radius map are firstly calculated from the fused binary map in the thresholding step. The skeletonization are realized by iteratively removing the outmost layer of pixels until unit thickness, as shown in **Figure 1(b)**. To acquire the radius map, the *distance field* is calculated [10], and the longest distance between pixels within the vessel and the nearest vascular boundary is considered as the vessel radius. The falsely acquired skeleton branches shorter than the vessel radius are discarded. To restore the vessel branches by FDHM, a circular mask with the size of the radius (RC) calculated above plus 15 μm, which is 3 pixels in this study ($Mask\ Radius = Calculated\ Radius + 3\ pixels$), is centered at each point along the skeleton, as shown in **Figure 1(b)**. Use this mask to take out the corresponding pixels of the HFE filtered image in the circle area. Keep the pixels of larger values than half of the maximum in the circle as the vessel segment and remove the rest. The entire vasculature is restored by repeating this process across all the skeleton points and adding all the segments up. The 3 extra pixels added on the RC mask radius ensures the full coverage of the target vessel. The full width at half maximum of a vessel's crossing profile is widely accepted as its segmentation [14]. The same idea is borrowed here with a circular implementation, and hence the vessel size distortions are corrected.

In the URG step, a modified region growing method is used to regain the connectivity between main vessels and adjacent small ones. Firstly, map the binary image from the last step onto the HFE filtered image, so that all the pixel values outside the foreground are set to 0, and thereby the vessel branches have clear boundaries. Secondly, the boundary layer pixels of a vessel branch are set as the initial seed points, and the URG only grows upwards to the pixels having larger values than the mean of the seed points. When a small vessel branch is originally connected to the adjacent large one, the pixels in between have larger values, so that the small vessel grows to the large one and regain the connectivity. If the small vessel is not originally connected to the large one, the pixels in between have smaller values and thus no growth. It should be noted that, no stop threshold is set for URG, which allows full growth for connectivity. The full growth without a stop threshold also causes vessel dilations, which would be corrected in the next step.

In the FDHM segmentation step, it is basically a repeat of the FDHM vessel restoration step but with the binary map from the URG step. Until here, the vessel segmentation is finished and all the distortions mentioned above are corrected.

## 2.2. 3D digital phantom and simulations

In order to evaluate the performance of the proposed algorithm, a 3D digital phantom is created to generate vascular images. The digital phantom is created from a 2D binary example vessel image from the open-source k-Wave Matlab Toolbox [15]. The skeleton and radius map are calculated from the image. For each pixel along the skeleton, a ball-shaped volume of 1s in 3D with the radius calculated from the radius map is centered at that pixel. By adding up all the balls across the skeleton, the 3D digital phantom is established from the 2D binary image, as shown in **Figure S 1 (a)**.

The simulation parameters of the light and transducer are designed according to our OR-PAM system [16]. The laser wavelength is set to 532 nm and the numerical aperture (NA) is set to 0.1. The central frequency of the transducer is set as 30MHz, and the one-way -6 dB bandwidth is 45 MHz. The unit distance between two adjacent pixels is set to 5 µm. For simplicity, both the light attenuation and acoustic attenuation are not accounted in the simulations [17].

The simulations run in 2D, where the vessel imaging follows a pattern of raster scanning, which scans the phantom vessel slice by slice, as shown in **Figure S 1 (b)**. The vessel cross-section is imaged pixel by pixel. All the detected B-scans are converted into MAP profiles and form the vessel image. The acquired MAP image is shown in **Figure 2 (b)**.

## 2.3. Performance evaluation

Accuracy (Acc), Sensitivity (Sn), Specificity (Sp) are three commonly used parameters to evaluate the segmentation performance. Acc provides the overall performance evaluation; Sn reflects the effectiveness of positive pixels detection; Sp specifies the effectiveness of negative pixels detection. The three metrics are defined as follows:

$$A_{cc} = \frac{TP + TN}{TP + FP + TN + FN} \tag{7}$$

$$S_n = \frac{TP}{TP + FN} \tag{8}$$

$$S_p = \frac{TN}{TN + FP} \tag{9}$$

where TP, TN, FP and FN are true positive, true negative, false positive and false negative, respectively.

The 2D projection of the digital phantom serves as the ground truth, as shown in **Figure 2(a)**. The proposed method is compared with two widely used segmentation techniques, OT and HF. The robustness of the proposed method is further tested with different scale ranges in comparison to HF. Three scale ranges are tested: $\sigma = [1\ 1.5], [1\ 3]$ and $[1\ 10]$

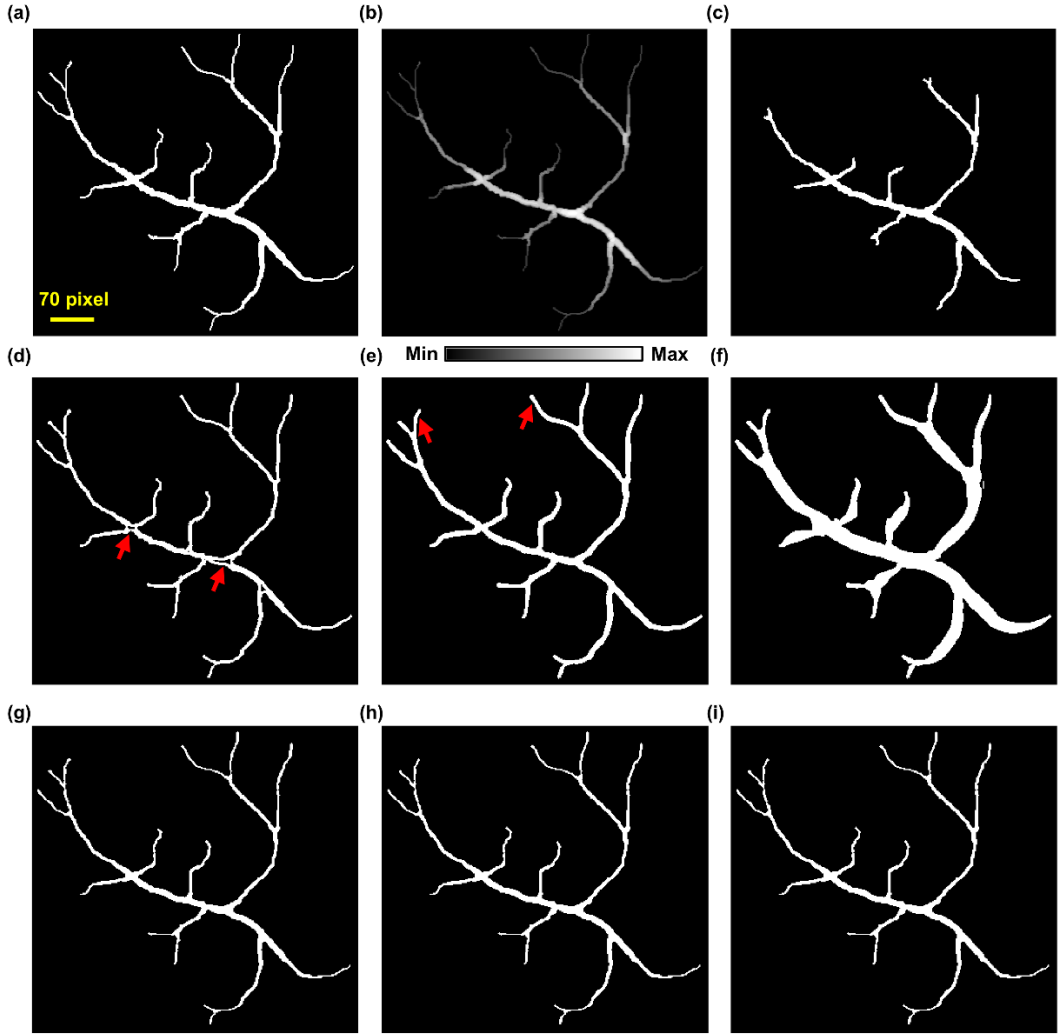

**Figure 2 Ground truth and MAP images of a digital vessel phantom and corresponding segmented results by the OT, HF, and proposed methods with different scale ranges.** (a) Ground truth. (b) MAP image. (c) Segmented result of the OT method. (d-f) Segmented results of the HF method with scale ranges of [1 1.5], [1 3] and [1 10], respectively; red arrows: under- and over-segmentation (g-i) Segmented results of the proposed method with scale ranges of [1 1.5], [1 3] and [1 10], respectively. MAP, maximum amplitude projection; OT, Otsu's thresholding; HF, Hessian filter.

## 2.4. OR-PAM system

To validate the proposed algorithm on *in vivo* data, we used a 2$^{nd}$ generation OR-PAM system described in our previous work [16]. The laser wavelength is 532 nm, and the NA of the objective is 0.1. The one-way -6 dB bandwidth of the transducer is 45MHz centering at 30MHz. The system provides a lateral resolution of 3 μm. The imaging is implemented by 2D raster scanning with a motorized stage. The step size is 5 μm.

## 2.5. *In vivo* vessel segmentation

*In vivo* experiments were performed on Swiss Webster mice (male; 8–9 weeks old; ~30 grams in weight). The protocol was approved by the Ethics Committee of Tsinghua University. Before imaging, the mouse was anesthetized with 1.5% isoflurane. A 1×1 mm² region of interest (ROI) was imaged both on the mouse cortex and ear vessels. The vascular images were then segmented by the proposed and HF methods with the scale ranges of [1 1.5], [1 3] and [1 10]. The Gaussian kernels of the three selected ranges can barely, mostly and overly cover all the target vascular sizes. Diameter distribution and VAF are calculated to evaluate the segmented results. VAF is defined as:

$$VAF = \frac{Vessel\ area}{Observed\ area} \quad (10)$$

where the vessel area is the area of the segmented foreground and the observed area is the area of ROI.

## 2.6. Longitudinal studies

The performances of the proposed method and HF method are compared on longitudinal studies. Longitudinal observations on the mouse ear vessel changes in response to epinephrine (EP) and sodium nitroprusside (SNP) were implemented. EP is an antihypotensive drug that causes vasoconstriction and SNP is an antihypertensive drug that leads to vasodilation. Prior to the drug injection, baseline images were recorded. The ROI of the mouse ear is 3×3 mm². A bolus of 4 μg EP solution and a bolus of 75 μg SNP solution were intravenously administered to different mice and continuous observations for 30 min started immediately after the drug injection. The diameter changes in representative vessels were quantitively analyzed.

## 3. Results

### 3.1. Simulation results

The OT, HF and proposed segmentation results on the digital phantom are demonstrated in **Figure 2**. The small vessel ends of weak signal are missed by the OT method, as shown in **Figure 2(c)**. The HF method with $\sigma =$ [1 1.5] fails to extract some branching points, as indicated by the red arrows in **Figure 2(d)**, demonstrating that small scale range causes under segmentations. The HF method with $\sigma =$ [1 3] shows dilation on small vessel ends, indicating the vessel enlargement effects from large Hessian kernels, as marked by the red arrows in **Figure 2(e)**. The HF method with $\sigma =$ [1 10] results in serious distortions in both size and shape, as shown in **Figure 2(f)**. The proposed method with the three different scale ranges shows identical results without apparent distortions, as shown in **Figure 2(g)-(i)**, and close to the ground truth (**Figure 2(a)**). The Acc, Sn and Sp values of different methods are listed in **Table 1**.

**Table 1** The accuracy (Acc), sensitivity (Sn) and specificity (Sp) of the OT, HF and proposed methods under different scale ranges.

|  | Acc | Sn | Sp |
|---|---|---|---|
| **OT** | 0.987 | 0.717 | 1 |
| **HF** $\sigma =$ [1 1.5] | 0.985 | 0.815 | 0.602 |
| **HF** $\sigma =$ [1 3] | 0.990 | 0.950 | 0.120 |
| **HF** $\sigma =$ [1 10] | 0.936 | 0.996 | 0.003 |
| **Proposed** $\sigma =$ [1 1.5] | **0.987** | **0.807** | **0.726** |
| **Proposed** $\sigma =$ [1 3] | **0.987** | **0.807** | **0.727** |
| **Proposed** $\sigma =$ [1 10] | **0.987** | **0.807** | **0.727** |

The OT method shows good outcome on Acc of 0.987 but poor result on Sn of 0.717, meaning that the positive pixels are less detected, which agrees well with the missing small vessel ends. The HF method is highly dependent on the scale range. The Acc increases from 0.985 to 0.99 when the scale range shifts from [1 1.5] to [1 3], by which the under segmentations are cleared. Acc decreases to 0.936 when the large scale range of [1 10] causes serious dilations. The Sp of the HF method drops from 0.602 to 0.12 and then to 0.003 when the scale range extends from [1 1.5] to [1 3] and [1 10], because the enlarged vessels reduce its detection of negative pixels. The proposed method with three different scale ranges shows identical results of 0.987 in Acc, 0.807 in Sn and 0.727 in Sp (0.726 at $\sigma =$ [1 1.5]), suggesting its high accuracy and robustness regardless of the different scale ranges. It should be noted that even if the OT method has high accuracy, the failure in small vessel extraction limits its use in complicated cases.

### 3.2. *In vivo* results

To better illustrate the effects of each image processing step, the intermediate images of the mouse cortex after major steps are presented in **Figure 3**. **Figure 3(a)** shows the original mouse cortex vessel image. The yellow arrows label the representative vessel structures that undergo distortions but are eventually corrected during the process. **Figure 3(b)** demonstrates the OT result, where evident missing vessels are observed, as indicated by the red arrow. **Figure 3(c)** shows the HF result. In this step, a scale range of [1 3] is carefully selected to match the majority of the vessel sizes. However, there are still under-segmented branching and crossing points, as well as over-segmented vessels as labelled by the red arrows. **Figure 3(d)** is the fused image of the OT and HF results, where the under segmentations from the HF step are corrected. The vessel disconnectivity caused by HF is labelled by the red arrows in the image. **Figure 3(e)** shows the image after the URG step, which regains the connectivity, but introduces new vessel dilations, as indicted by the red arrows. The final result after the FDHM segmentation is shown in **Figure 3(f)**, where all the previously distorted structures are corrected, as marked by the green arrows. **Figure 3(g)** shows the signal profiles of the original image, the results of the OT, HF and proposed methods along the dashed line in **Figure 3(a)**. The small vessel of weak signal in the middle is successfully extracted by both the HF and proposed methods but is missed by the OT method. The large vessel on the left is extracted at the full width at half maximum (FWHM) by the proposed method, but at larger size by the OT and HF methods. On this vessel, the proposed method shows better accuracy over the OT and HF methods. Furthermore, the vessel on the right is also segmented at FWHM. According to **Figure 3(g)**, the proposed method applies different thresholds to different vessels, which at best preserves every single vessel's size. In short, the proposed method is able to provide accurate and robust results on the *in vivo* data.

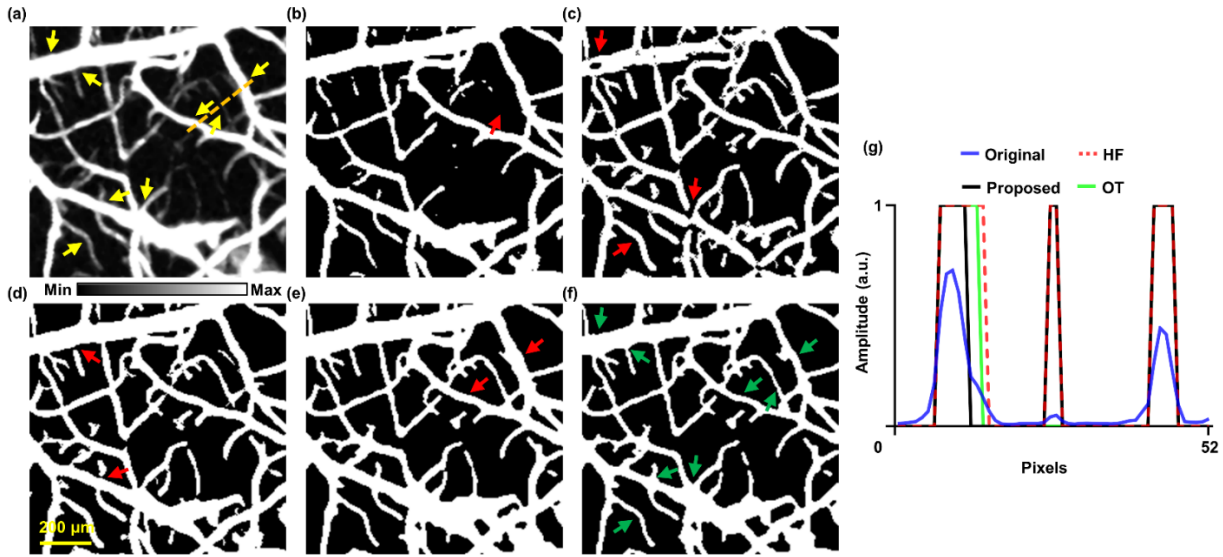

**Figure 3 Illustration of the intermediate images after major steps of the proposed method and vessel profiles of the original, OT, HF and proposed methods.** (a) Original mouse cortex vessel image; yellow arrows: vessel structures experiencing distortions during the segmentation process. (b) Result of the OT method; red arrow: under-segmented vessel. (c) Result of the HF method; red arrow: under-segmented branching and crossing points and over-segmented vessel branch. (d) Fusion image of the OT and HF results; red arrows: vessel disconnectivity caused by HF. (e) Results after FDHM vessel restoration and URG; red arrows: dilated vessel branch caused by URG. (f) Final result after FDHM segmentation; green arrows: corrected vessel structures. OT, Otsu's thresholding; HF, Hessian filter; FDHM, full diameter at half maximum; URG, upwards region growing.

The impacts of different scale ranges on the proposed method and the HF method are evaluated on *in vivo* vasculature. The segmentations with different scale ranges on the mouse cortex vasculature are shown in **Figure 4**. The original vessel image is shown in **Figure 3(a)**. The results of the proposed method with scale ranges of [1 1.5], [1 3] and [1 10] are shown in **Figure 4(a)**, **Figure 4(c)** and **Figure 4(e)**, respectively. The results of the HF method with the three scale ranges are shown in **Figure 4(b)**, **Figure 4(d)** and **Figure 4(f)**, respectively. **Figure 4(g)-(l)** illustrate the corresponding diameter distributions of **Figure 4(a)-(f)**. According to **Figure 4(g)**, **Figure 4(i)** and **Figure 4(k)**, despite the slight differences among the three diameter distributions, the general distribution trends of the proposed method are similar to one another. On the contrary, the diameter distribution trends under different scale ranges of the HF method are completely different from one another, as shown in **Figure 4(h)**, **Figure 4(j)** and **Figure 4(l)**. The vessel sizes under $\sigma = $ [1 1.5] are mostly concentrated at 10-30 μm (**Figure 4(h)**), because under segmentation turns the large vessels into multiple small ones, as marked by the red arrow in **Figure 4(b)**. With the wider scale range of [1 3], the vessel sizes of 40-60 μm appears as the under segmentations are mitigated with larger Hessian kernels (**Figure 4(d)**). The scale range of [1 10] dilates all the vessels (**Figure 4(f)**), which shifts the entire diameter distribution towards the higher end, as shown in **Figure 4(l)**. The segmentation of mouse ear vessels and corresponding diameter distributions are shown in **Figure S 2**, where similar results are demonstrated. To be noted, large Gaussian kernel creates or restores vessel continuity [8], which could be the reason that leads to the slight structural differences in **Figure 4(a)**, **Figure 4(c)** and **Figure 4(e)**, as indicated by the red arrows. The consistent diameter distributions under different scale ranges in the proposed method validates its robustness *in vivo*.

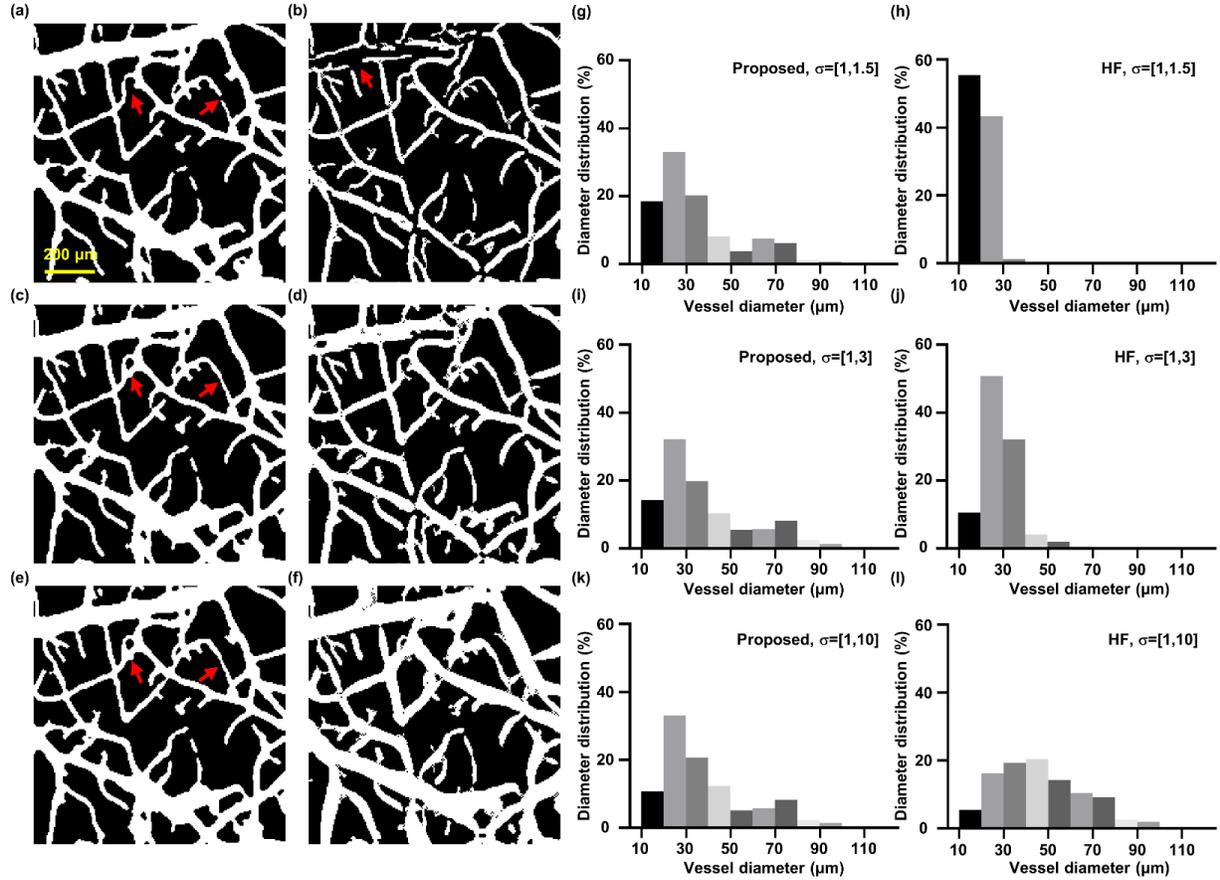

**Figure 4 Vessel segmentation results on mouse cortex vessels of the proposed method and the HF method with different scale ranges, and corresponding vessel diameter distributions**. (a) Result of the proposed method, $\sigma =$ [$\mathbf{1\ 1.5}$]; red arrow: structural differences to (c) and (e). (b) Result of the HF method, $\sigma =$ [$\mathbf{1\ 1.5}$]; red arrow: under segmented vessel structure. (c) Result of the proposed method, $\sigma =$ [$\mathbf{1\ 3}$]; red arrow: structural differences to (a) and (e). (d) Result of the HF method, $\sigma =$ [$\mathbf{1\ 3}$]. (e) Result of the proposed method, $\sigma =$ [$\mathbf{1\ 10}$]; red arrow: structural differences to (a) and (c). (f) Result of the HF method, $\sigma =$ [$\mathbf{1\ 3}$]. (g-l) Corresponding diameter distributions in (a-f). HF, Hessian filter.

The VAFs of the proposed method and HF method on the mouse cortex vasculature are listed in **Table 2**. The VAFs from the proposed method have differences of ~0.01 between each two adjacent scale ranges, while those from the HF method show a difference of ~0.1. The VAF errors caused by the three scale ranges of the proposed method are 10 times smaller than those in the HF method, which further proves the robustness of the proposed method.

**Table 2 The VAF of the segmented vessels from the proposed and HF method in different scale ranges**

| Scale range | Proposed | HF |
|---|---|---|
| [1 1.5] | 0.330 | 0.224 |
| [1 3] | 0.344 | 0.332 |
| [1 10] | 0.353 | 0.445 |

### 3.3. Longitudinal results

The performance of the proposed method on quantitative analysis of longitudinal applications are validated. Diameter changes on the mouse ear vasculature in response to epinephrine administration are shown in **Figure 5**. The original OR-PAM MAP image is shown in **Figure 5(a)**. A subset ROI of a branching structure is segmented by both the proposed and HF methods with a scale range of [1 5]. The diameter mean of an artery, labelled in blue in **Figure 5(a)**, is calculated at different time points. The segmented results by the proposed method at baseline (BL), 3 min after injection (+3 min) and 15 min after injection (+15 min) are shown in **Figure 5(b-d)**. Evident vasoconstriction can be observed at +3 min, as marked by the yellow arrows. The segmented results by the HF method at BL, +3 min and +15 min are shown in **Figure 5(e-g)**. The vessel diameter changes are normalized to their baseline values and the relative changes are demonstrated in **Figure 5(h)**. An apparent stronger vasoconstriction of - 50.3 % is detected by the proposed method compared to that of - 24.6 % in the HF method at +3 min. The vasoconstrictions presented by the proposed and HF methods at +15 min are - 37.5 % and - 9.6 %, respectively. The vasoconstrictive response after epinephrine injection followed by a progressive recovery agrees well with previous studies [18]. In this longitudinal study, a ~ 2 times vasoconstriction is detected by the proposed method compared to the HF method. Moreover, the quantitative analysis errors introduced by different techniques can seriously affect the final results.

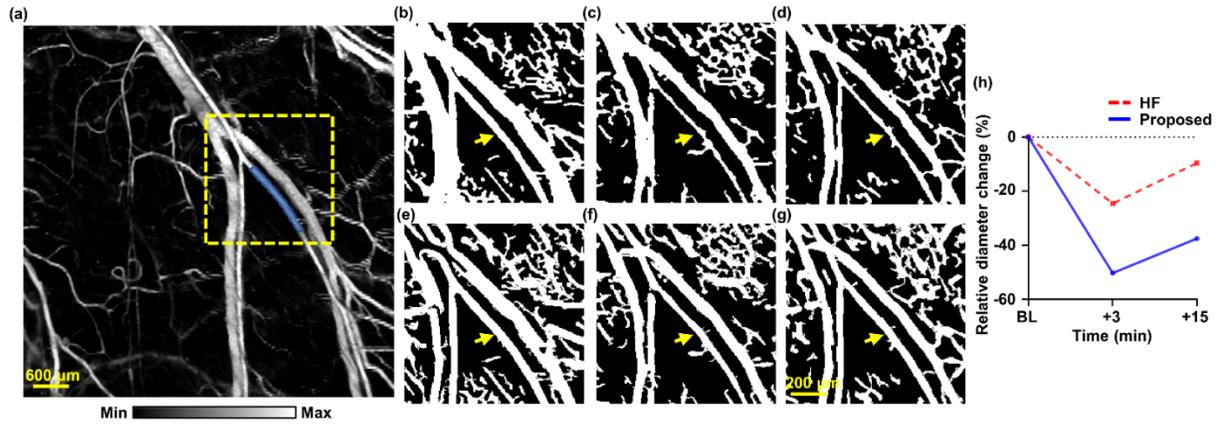

**Figure 5 Quantitative analysis of the vessel diameter changes in response to EP injection**. (a) Original OR-PAM MAP image; dashed box: a subset ROI to be segmented, blue area: target artery segment for diameter analysis. (b)-(d) Segmented results from the proposed method at BL, +3 min and +15 min; yellow arrow: target artery. (e)-(g) Segmented results from the HF method at BL, +3 min and +15 min; yellow arrow: target artery. (h) Relative diameter change calculated by the proposed and HF methods. EP, epinephrine; OR-PAM, optical-resolution photoacoustic microscopy; MAP, maximum amplitude projection; ROI, region of interest; HF, Hessian filter; BL, baseline

Another longitudinal study of the vessel size changes induced by SNP administration on a mouse ear is implemented. The original OR-PAM image is shown in **Figure 6(a)**. A subset ROI is segmented by both the proposed method and the HF method with a scale range of [1 5], as marked in the dashed box in **Figure 6(a)**. The diameters within a selected artery are averaged in the longitudinal analysis, as labelled in blue in **Figure 6(a)**. **Figure 6(b)-(d)** demonstrate the segmented results by the proposed method at BL, +3 min and +15 min, respectively. Vasodilation can be observed after SNP injection at +3 min, as marked by the yellow arrows. The segmented results by the HF method at BL, +3 min and +15 min are shown in **Figure 6(e)-(g)**. The relative diameter changes calculated by the two methods are presented in **Figure 6(h)**. The vessel diameter presented by the proposed method has an increase of 18.8 % at +3 min, and drops down to -0.6 % at +15 min. The vessel segmented by the HF method shows a vasodilation of 2.8 % at + 3min, and returns to – 1.1 % at + 15min. In this case, a much stronger vasodilation is detected by the proposed method after the SNP injection, compared to that by the HF method. To be noted, an arterial diameter increase of 2.8 % to external stimuli can be negligible as arteries exhibit fluctuations under normal states [19]. Hence, with the results from the HF method of the scale range of [1 5], we cannot conclude whether there is vasodilation in response to SNP.

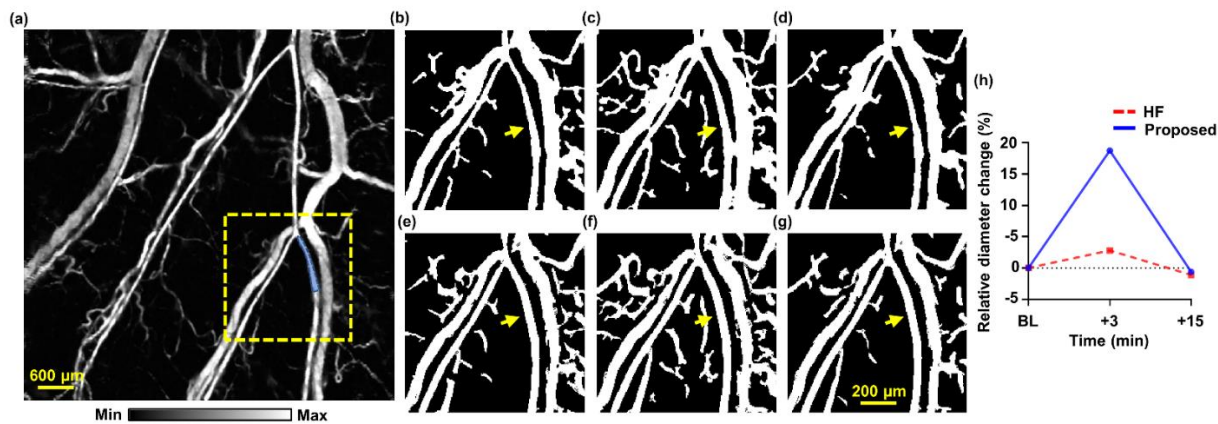

**Figure 6 Quantitative analysis of the vessel diameter changes in response to SNP injection**. (a) Original OR-PAM MAP image; dashed box: a subset ROI to be segmented, blue area: target artery segment for diameter analysis. (b)-(d) Segmented results from the proposed method at BL, +3 min and +15 min; yellow arrow: target artery. (e)-(g) Segmented results from the HF method at BL, +3 min and +15 min; yellow arrow: target artery. (h) Relative diameter change calculated by the proposed and HF methods. SNP, sodium nitroprusside; OR-PAM, optical-resolution photoacoustic microscopy; MAP, maximum amplitude projection; ROI, region of interest; HF, Hessian filter; BL, baseline.

The results from both longitudinal studies collectively show that the proposed method is more sensitive to subtle vessel changes than the HF method. The HF method with a fixed scale range either enlarges the constricted vessels or thins the dilated ones because the Hessian kernels only extract the vessels of matching sizes, which can explain its insensitivity to the subtle vascular changes. On the other hand, segmenting at the half maximum in the proposed method would closely capture the real vascular changes.

## 4. Conclusion and Discussion

In this work, we have developed a Hessian filter-assisted, adaptive thresholding vessel segmentation algorithm. Its performance is quantitively evaluated on a 3D custom-designed digital vessel phantom, demonstrating strong robustness and high accuracy. Further tests on *in vivo* vascular images show promising consistency in vessel structure preservation regardless of the different scale ranges. We have also compared the proposed algorithm with the HF method on two longitudinal studies of vascular changes in response to different blood pressure regulating drugs. The results from both studies show much higher sensitivity of the proposed algorithm in capturing subtle vessel changes over the HF method.

Different from those single-value thresholding methods or adaptive thresholding methods [20-22], the basic idea of the proposed algorithm is to apply different thresholds to each point along the vessel skeleton, in order to better preserve the vascular structures at every point. A fixed threshold applied to different vessels would either overly segment the vessels of larger values or insufficiently segment the vessels of smaller values, both leading to distortions. The strategy proposed here have solved this problem.

The role of the HF step and skeletonization step in the algorithm is to enhance the vessels of weak signal and determine their positions. Without the enhancement, the small vessels can be possibly missed if only gray-scale information is applied [6]. Moreover, if two adjacent vessels are accidentally merged by a large scale range, the proposed method can separate them away as long as the half maximum value is able to differentiate them, as demonstrated by the red arrows in **Figure S 2(e) and (f)**. This also contributes to the robustness of the algorithm.

Despite the resilience to the impacts of different scale ranges, the proposed algorithm is still affected by different Gaussian kernels. Large Gaussian kernels create or restore continuity on broken vessels, which otherwise remain unconnected, as demonstrated by the red arrows in **Figure 4(a)**, **Figure** 4**(c)** and **Figure** 4**(e)**. Therefore, a proper scale range matching the majority of the vessels in an image is still preferred in the proposed method.

During the process of the proposed algorithm, multiple parameters are selected empirically by conducting multiple trials. In the HFE step, the parameter *a* controlling the offset is set to 2 and the parameter *b* controlling the high-frequency emphasis is set to 3. An overly large ratio of *b* to *a* would enlarge small vessels and thin large ones, thereby affecting the segmented results. The clipping limit controlling the local contrast enhancement in the CLAHE step is set to 0.005. A higher limit would result in more local contrast, where enhanced background noise could be recognized as vessel structures. The sizes of the two Gaussian filters deciding the thresholds to binarize the HF filtered images in the adaptive thresholding step are set to 3 and 11 pixels. These two factors decide how weak the signal is to be included as vessel structures. These parameters need to be prudently selected, which otherwise would induce errors to the final results. Therefore, fully automatic segmentation strategies are in particular needs. Novel imaging processing techniques based on neural networks have shown promising potentials on fully automatic vessel segmentation [23, 24]. We expect that the development of such techniques may further push the limits on fully automatic vessel segmentation.

**Acknowledgement**

This work was partially supported by the National Natural Science Foundation of China (61871251, 62027901 and 82071925).

**Declaration of Competing Interest**

The authors declare that they have no known competing financial interests or personal relationships that could have appeared to influence the work reported in this paper.

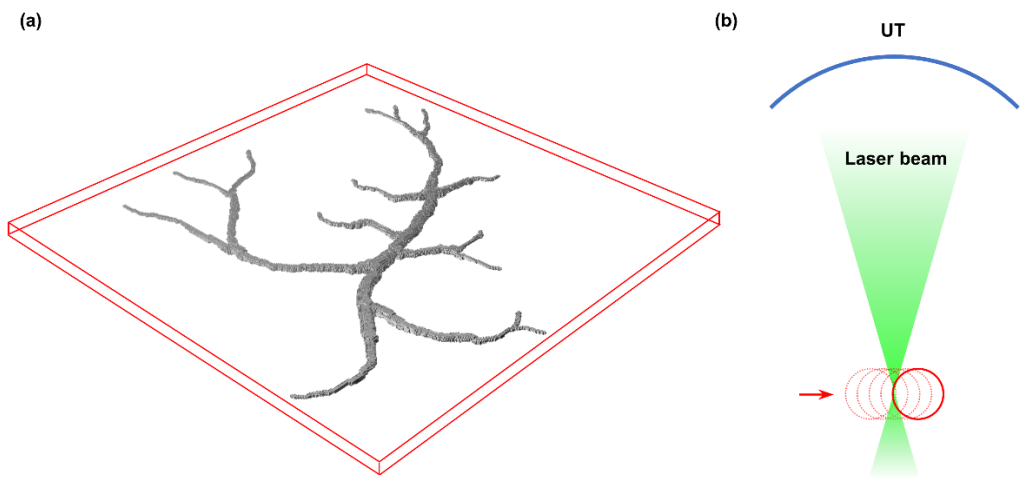

**Figure S 1 The established 3D digital vessel phantom and the schematic of 2D scanning simulations.** (a) 3D digital vessel phantom. (b) The schematic of 2D vessel scanning setup in the simulations; red circle: vessel cross-section; red arrow: scanning direction. UT: ultrasound transducer.

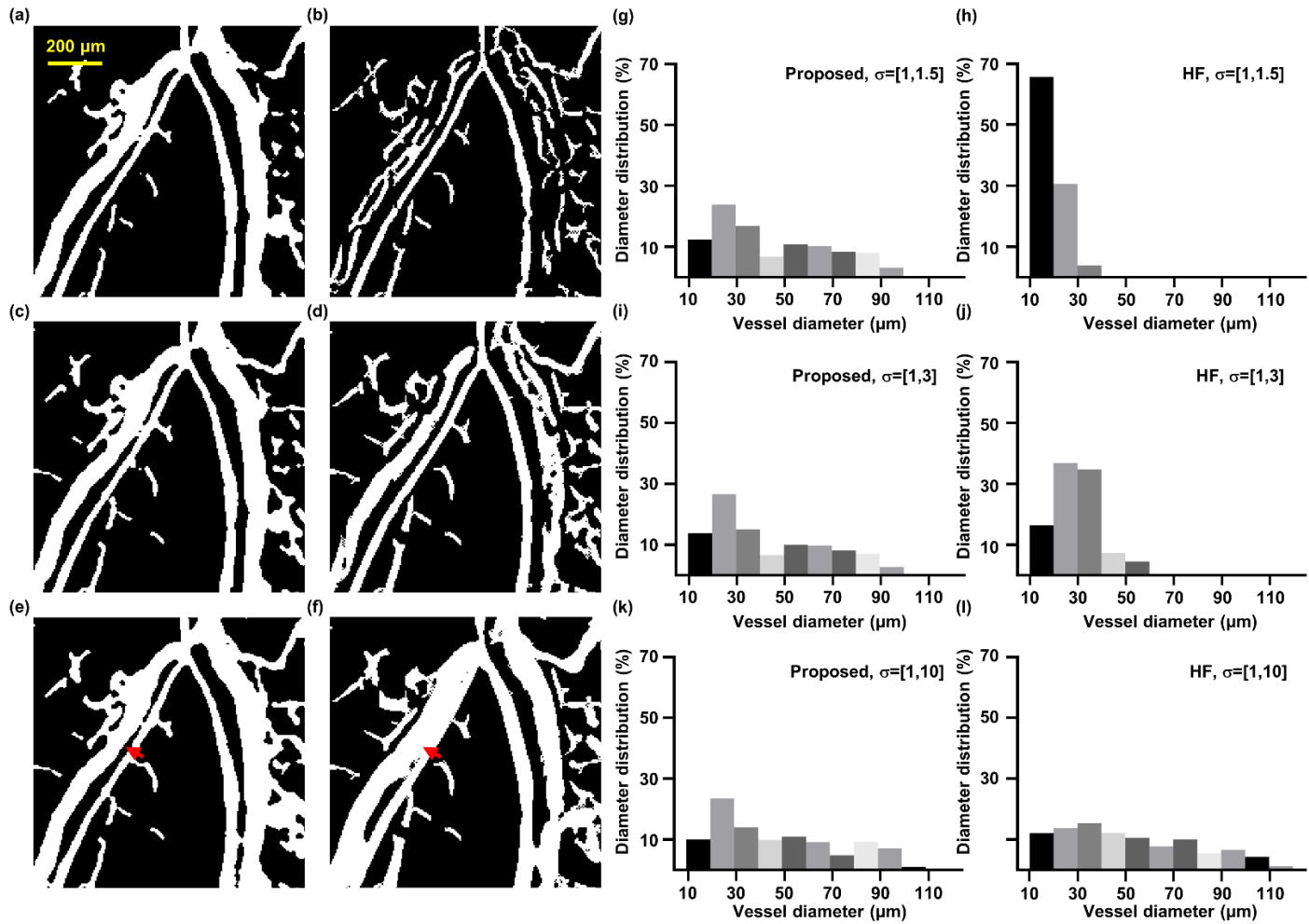

**Figure S 2 Vessel segmentation results on mouse ear vessels of the proposed method and HF method with different scale ranges, and corresponding vessel diameter distributions**. (a) Result of the proposed method, $\sigma = [1\ 1.5]$. (b) Result of the HF method, $\sigma = [1\ 1.5]$. (c) Result of the proposed method, $\sigma = [1\ 3]$. (d) Result of the HF method, $\sigma = [1\ 3]$. (e) Result of the proposed method, $\sigma = [1\ 10]$; red arrows: separated vessels. (f) Result of the HF method, $\sigma = [1\ 10]$; red arrow: merged vessel. (g-l) Corresponding diameter distributions in (a-f). HF, Hessian filter.